\documentclass[preprint,12pt]{elsarticle}

\usepackage{amssymb}
\usepackage[T1]{fontenc}

\date{}
\begin{document}

\begin{frontmatter}



\title{Examining composition-dependent radiation response in AlGaN}


\author[inst1]{Miaomiao Jin\fnref{label2}}
\fntext[label2]{Corresponding author: mmjin@psu.edu}

\affiliation[inst1]{organization={Department of Nuclear Engineering, Pennsylvania State University},
            addressline={205 Hallowell Building}, 
            city={University Park},
            postcode={16802}, 
            state={PA},
            country={US}}

\author[inst1]{Farshid Reza}
\author[inst1]{Alexander Hauck}
\author[inst1]{Mahjabin Mahfuz}
\author[inst1]{Xing Wang}
\author[inst3]{Rongming Chu}
\author[inst2]{Blair Tuttle}
 
\affiliation[inst3]{organization={Department of Electrical Engineering, Pennsylvania State University},
            addressline={N-237 Millennium Science Complex}, 
            city={University Park},
            postcode={16802}, 
            state={PA},
            country={US}}
            
\affiliation[inst2]{organization={Department of Physics, Penn State Behrend},
            addressline={116 Witkowski}, 
            city={Erie},
            postcode={16563}, 
            state={PA},
            country={US}}

\begin{abstract}
Al$_x$Ga$_{1-x}$N materials have become increasingly important for electronics in radiation environments due to their robust properties. In this work, we aim to investigate the atomistic mechanisms of radiation-induced damage in AlGaN compounds, providing insights that bridge the gap between high-length-scale experimental data and detailed atomic-level processes. Through extensive molecular dynamics simulations, we reveal the compositional dependence of radiation-induced defect production in Al$_x$Ga$_{1-x}$N systems with $x$ ranging from 0 to 1. The damage accumulation characteristics observed in our simulations align notably well with available experimental data at temperatures up to room temperature. Our findings indicate that alloy composition significantly influences defect production and microstructural evolution, including the formation of dislocation loops and defect clusters. Specifically, with increasing Al content, defect production from individual recoil events is reduced; however, extended interstitial defects are more likely to form considering cumulative effects, leading to enhanced damage at high doses. Among the compositions studied, we find that 25\% Al content leads to the least overall radiation damage, suggesting an optimal alloying strategy for mitigating radiation effects. These findings underscore the interplay between defect formation, dynamic annealing, and cascade effects, offering insights for optimizing AlGaN materials for radiation resistance in practical applications.

\end{abstract}




\end{frontmatter}


\section{Introduction}
The study of radiation damage in GaN-based materials systems becomes increasingly crucial, particularly given the widespread use of these materials in optoelectronics, high-electron mobility transistors (HEMTs), and power devices. Their wide bandgap, high electron mobility, and enhanced radiation tolerance compared to Si-based electronics make understanding the radiation effects essential for applications in space, nuclear, and high-radiation environments \cite{weaver2016radiation,pearton2015ionizing}. This irradiation resistance was attributed to GaN's high displacement energy and strong defect annealing \cite{pearton2015ionizing}. However, despite its inherent resistance to radiation, GaN heterostructures are not immune to irradiation degradation. Ion irradiation serves as a key tool to investigate radiation response, including electrical degradation and structural damage. When subjected to high-energy ions, Al$_x$Ga$_{1-x}$N materials exhibit various defects, including point defects, defect clusters, dislocations, stacking faults, and in extreme cases, amorphization and phase decomposition \cite{liu1998ion,kucheyev2000ion,wang2002defect,pkagowska2011rbs,islam2019heavy}. With swift heavy ions, latent tracks are found due to strong ionization \cite{lei2018degradation,mahfuz2024microstructural}. The formation of these defects can serve as trapping and scattering centers for charge carriers \cite{weaver2012displacement, karmarkar2004proton}, resulting in reduced sheet carrier density and carrier mobility. With defects positioned near the two-dimensional gas (2DEG) channel region, degradation of performance has been frequently reported in GaN-based devices \cite{hwang2014effect,pearton2015ionizing,kim2012proton,luo2001dc,puzyrev2011radiation,islam2019heavy}. Although most of the studies attribute the degradation to displacement damage which leads to the creation of deep-level compensating lattice defects \cite{crawford1959nature},  significant linear energy transfer due to electronic interaction can lead to pronounced single event effects, where a single passage of ion can disrupt device operation \cite{rostewitz2013single,kuboyama2011single,bazzoli2007see}.  Furthermore, the formation of latent tracks at high electronic energy loss regime can also lead to degradation of electrical characteristics of HEMTs such as gate leakage current, threshold voltage and saturation drain current \cite{hu2018degradation,lei2018degradation}.

The displacement damage process starts with sub-femtosecond level collisions, which makes it impractical to directly locate the defects from the primary damage stage (picoseconds) and monitor the short timescale defect evolution with experimental techniques. The difficulty in locating defects and monitoring their evolution highlights the necessity for complementary simulation methodologies to unravel the atomistic nuances of radiation effects in these materials. Indeed, computational studies have been frequently performed in GaN to examine the defects induced by energetic recoils. These efforts yielded rich discussions on threshold displacement energy (TDE) and the properties of resultant defects. Notably, the exact values of TDE and the consequent defects produced from near-threshold recoils vary between different methods and force fields \cite{nord2003molecular,xiao2009threshold,song2023neural}. For example, Nord et al. suggested 18$\pm$1 eV for Ga and 22$\pm$1 eV for N for minimal TDE based on classical molecular dynamics (MD) simulations \cite{nord2003molecular}, while Xiao et al. evaluated the minimal to be 39 eV for Ga and 17 eV for N using ab-initio MD simulations \cite{xiao2009threshold}. However, recently, by screening a significant amount of crystalline directions, Hauck et al. noted 22$\pm$1 eV for Ga and 11$\pm$1 eV for N from ab-initio MD simulations, in close agreement with experimental values \cite{hauck2024atomic,ionascut2002radiation,look1997defect}. The defects formed after the initial knock-on events are generally point defects and small defect complexes \cite{hauck2024atomic,xiao2009threshold}.  There is a significant number of studies discussing the characteristics of point defects and their connection with the degradation of device electrical metrics; these defects can be readily created under thermal and fabrication conditions. However, the role of defect complexes can not be neglected, particularly considering displacement radiation damage. For example, it was suggested that the transformation of a $\mathrm{V_{Ga}}$ to a Ga-N divacancy due to displacement damage can cause a negative shift in the threshold voltage, leading to an increase in the $1/f$ noise measurements of proton-irradiated GaN/AlGaN HEMTs \cite{puzyrev2011radiation}.  With high-energy (10 keV) recoils in GaN, MD simulations of single radiation damage cascades have already demonstrated a significant amount of defect clustering \cite{he2020primary}. 

Although there have been numerous irradiation studies on GaN, concerning the electronic properties and structural changes under irradiation \cite{pearton2015ionizing}, only limited studies exist for the AlGaN alloy systems. Using fast neutron irradiation with fluences of 1–3$\times10^{15}$cm$^{-2}$ on Al$_x$Ga$_{1-x}$N/AlN/GaN HEMT structures at different Al concentrations (20\%, 30\%, and 40\%),  a decrease in both 2DEG concentration and accumulation capacitance was found. The strongest effect was reported for the highest Al content:  the introduction rate of trap density is highest in 40\% Al sample based on the threshold voltage shift. It implies that these radiation-induced traps are located in the AlGaN barrier instead of the GaN buffer or the interface \cite{polyakov2012comparison}. Not only the initial trap density from the quality of sample fabrication using different Al concentrations \cite{polyakov2012comparison,pearton2015ionizing}, but the radiation-induced defect evolution considering the variation of composition in AlGaN could contribute to the difference. With the alloying of Al in GaN, Faye et al. considered the full compositional range of Al$_x$Ga$_{1-x}$N subject to 200 keV Ar irradiation and noted that Rutherford backscattering spectrometry/channeling (RBS/C) reveals a higher damage level with high Al concentration at high fluences, which is against the common belief that AlN is more radiation resistant \cite{faye2016mechanisms}. They ascribed this to the formation of stable extended defects. A later study \cite{faye2018crystal} found that at low doses, AlN incorporation enhances radiation resistance, while for high doses, damage level was enhanced, which was attributed to nitrogen deficiency. However, to the authors' knowledge, a fundamental understanding towards the radiation damage processes in these alloy systems has not been elucidated.   

An understanding of the interaction between radiation-induced defects in these systems would assist in the current efforts to enhance the radiation tolerance of AlGaN through material engineering strategies such as alloying. The damage build-up processes represent complex defect dynamics, particularly for the ternary Al$_x$Ga$_{1-x}$N systems. Therefore, in this work, we aim to provide a systematic study of the characteristics of radiation-induced defects in composition-dependent AlGaN systems. To accomplish this, we utilize classical MD to extensively evaluate the defect generation and the interactions among defects during ion irradiation. The low-temperature ion irradiation experiments are used for validation purposes. Defect analysis reveals that the formation of defect clusters and extended defects can be from the dynamic annealing during the thermal spike processes in systems containing pre-existing defects, even without long-range thermally activated diffusion.  The excellent comparisons with available experimental data allow a fundamental understanding of microscopic mechanisms of irradiation damage in AlGaN systems.

\section{Methods}

We conducted classical MD simulations as implemented in the LAMMPS package \cite{plimpton2007lammps} to investigate radiation damage in Al$_x$Ga$_{1-x}$N systems, with $x$ being 0.0, 0.25, 0.50, 0.75, and 1.0. The simulation cell used to simulate the radiation damage cascade is in a wurtzite structure consisting of 92,160 - 1,034,880 atoms (box size varies based on the recoil energy). Periodic boundary conditions are applied along all three axial directions. Atomic interactions are described by a Stillinger-Weber (SW) potential from Zhou et al. \cite{zhou2013molecular}. This potential has been demonstrated to accurately predict the stable phases, cohesive energies, and lattice constants for the relevant phases and has also been used to predict interfacial thermal transport in Al/GaN \cite{zhou2013molecular,zhou2013relationship}. Given its formalism, SW is exceptionally hard for short-range interactions (see supplementary materials (SM) for visual demonstration), which would be problematic for describing the collisions from high-energy recoils. Instead, Ziegler-Biersack-Littmark (ZBL) repulsive potential \cite{ziegler1985stopping}, a screened Coulomb potential, is used for the short-range interactions. This is achieved via smoothly splining the two potentials over 0.7 {\AA} to 0.8 \AA. To prevent nonphysical results, we have modified the original SW formalism to flatten the short-range uptick curves with a smooth transition function. Then, the potential ``overlay'' functionality within LAMMPS is used to achieve continuity in energy and force. Detailed pair interaction energies for the modified SW and ZBL potentials can be found in the SM. 

With relaxed simulation cells (under isothermal-isobaric ensemble with zero external pressure at 15 K or 300 K for 100 picoseconds), to initiate a damage cascade, we randomly select one Ga/Al atom as the primary knock-on atom (PKA) and assign this recoil a specific kinetic energy directed along a crystalline direction. In addition, an adaptive time-stepping method is employed, limiting atomic movements to 0.05 {\AA} per timestep for enhanced accuracy in capturing the collision processes. To mimic an environment that drains the excess energy introduced by the PKA, a thermal bath at a prescribed ambient temperature is achieved by applying a temperature-rescaling thermostat to the atoms in the outermost layer of the cell, with a thickness of one lattice constant. To ensure the damage cascade induced by the PKA is fully contained within the cell without crossing the boundary, atoms are shifted such that the PKA is initially located at the center of the cell. For each cascade process, the system is sufficiently annealed to the ambient temperature over approximately 30 picoseconds.

Based on this methodology, we conducted two types of MD simulations to examine the composition-dependent radiation damage. The first type is the single cascade simulation, where we explore the effects of initiating the PKA along random directions and particular crystalline directions ($\mathbf{a}$, $\mathbf{c}$, and $\mathbf{m}$ directions). This choice is informed by previous experimental studies on epitaxial GaN layers, which demonstrate different radiation responses depending on the crystalline orientation with respect to the beam direction: the damage saturation level at high ion fluence is lower in $\mathbf{a}$-plane layers compared to $\mathbf{m}$- and $\mathbf{c}$-plane samples \cite{lorenz2017implantation,mendes2019measuring}. Given this observation, we will examine how defect formation differs, considering variations in PKA direction and kinetic energy. To account for statistical variation, ten independent simulations are averaged for each scenario. The second type is multi-cascade simulation. To consider the cumulative effects of radiation damage, we adopt the methodology described in \cite{jin2020achieving} for simulating overlapping cascades to account for the interaction of cascades with pre-existing defects. This procedure involves randomly introducing consecutive PKAs into the simulation cell. Each PKA is assigned 2 keV of kinetic energy with a randomly chosen direction, thereby allowing us to simulate a series of radiation cascade events that accumulate and potentially overlap. By increasing the number of PKAs, one may obtain $\sim$1 dpa dose level according to the Norgett-Robinson-Torrens (NRT) model \cite{norgett1975proposed,jin2020achieving}. It effectively provides a detailed view of the cascade dynamics and defect formation under various initial conditions. Six independent simulations are averaged for each composition to ensure the statistical significance of conclusions. Note that the consequent dose rate is orders of magnitude higher than that in experimental conditions due to the limited MD timescale. Despite this, the methodology allows for examining defect-defect interactions and exploring defect dynamics within temperature regimes where microstructural changes are primarily driven by damage cascades. This approach is particularly relevant for comparison with low-temperature experiments (such as cryogenic temperatures), where long-range thermal diffusion is minimal.

For analysis purposes, atomic configurations of the system at increasing timesteps/doses are recorded. Results presented in the figures are based on 300 K simulations, and additional results at 15 K are provided in the SM. With the configurations, defects and defective structures are identified using the Wigner-Seitz cell method \cite{nordlund1998defect} and common neighbor analysis, facilitated by the OVITO software package \cite{stukowski2009visualization}. Defect clustering is analyzed based on their distances, with a cutoff of 3.5 {\AA}. Note that antisite defects are excluded from clustering analysis to avoid ambiguity in defining cluster size. For strain analysis, the atomic configurations are dynamically relaxed under isothermal-isobaric conditions with zero external pressure. The time-averaged dimensions are recorded to evaluate strain as a function of radiation dose.

\section{Results}

\subsection{Simulation of Primary Radiation Damage}

\begin{figure}[ht]
    \centering
    \includegraphics[width=0.99\textwidth]{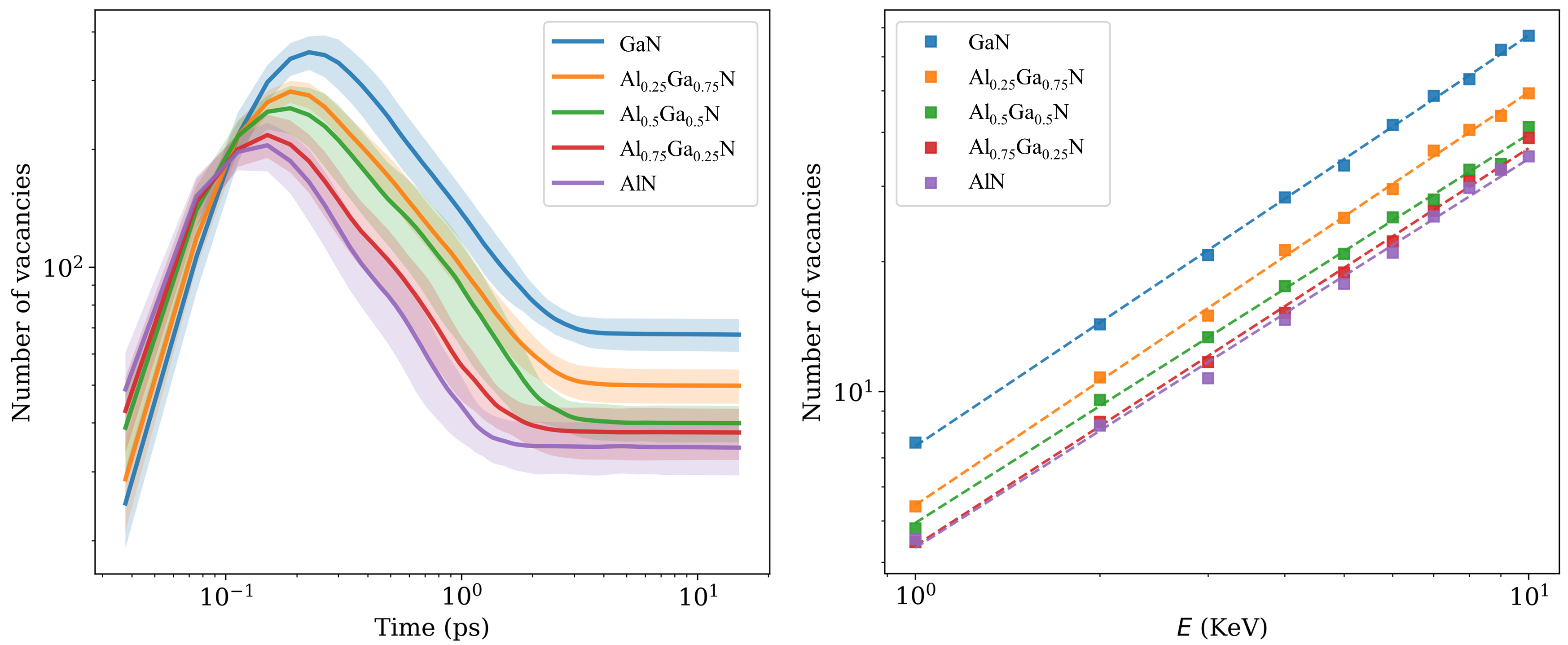}
    \caption{(a) Number of vacancies versus time for 10 keV Ga PKA induced damage cascades and (b) number of vacancies produced versus Ga PKA energies within 1-10 keV in 300 K Al$_x$Ga$_{1-x}$N systems. The shaded areas in (a) indicate the standard deviation. }
    \label{fig:single_defects_energy}
\end{figure}

To understand the defect production, the number of vacancies is identified as a function of time. With a 10 keV Ga PKA, Figure \ref{fig:single_defects_energy}a exhibits a generic pattern for the primary radiation damage \cite{bacon2000primary}: regardless of the composition, the number of defects starts to increase drastically over sub-picoseconds, followed by a thermal spike phase and subsequent annealing. The final number of defects is significantly lower than the peak damage level and varies depending on the alloy composition. We note several observations based on the comparison. i) The duration of the damage cascade is influenced by the composition, with pure GaN exhibiting the longest cascade lifetime compared to other compositions. Overall, the cascade lifetime decreases as the Al content increases. ii) The time at which peak damage occurs decreases with increasing Al content, indicating that the presence of Al accelerates the initial damage phase. iii) The peak number of defects reduces with increasing Al content, suggesting that higher Al concentrations mitigate the extent of the damage cascade. iv) After the annealing of the cascade, the final number of defects decreases with increasing Al content, although the differences become smaller at higher Al concentrations. This temporal evolution of defect formation is linked to the thermal transport efficiency and amorphization during the thermal spike phase, which can be partially understood by examining thermal diffusivity and melting temperature \cite{averback1994atomic}. Although it is unfeasible to know the exact values of these thermal properties as they depend on the local temperature, pressure, and structure during cascade-induced transients \cite{zarkadoula2013nature}, to a first approximation, the high thermal conductivity in AlN \cite{daly2002optical} allows for more efficient heat dissipation \cite{mahfuz2024microstructural}, thereby reducing the window for defect formation. 

The final number of defects following cascade annealing is crucial for assessing the stable defects that are likely to undergo evolution. Hence, we analyze defect production as a function of the PKA kinetic energy. As illustrated in Figure \ref{fig:single_defects_energy}b, the number of defects increases with PKA energy. With increasing Al content, the number of vacancies is consistently reduced across the energy range up to 10 keV. As a comparison for the GaN, previous studies using another interatomic potential (Tersoff) have reported varying numbers of defect production for a 10 keV Ga damage cascade, with one study noting 51.3 Frenkel pairs \cite{he2020primary} and another reporting 46.7 vacancies \cite{nord2003molecular}, both are lower than the current value 67.3$\pm$6.5. These differences are likely attributable to the capabilities of the interatomic potentials in the medium to near-equilibrium range, as the short-range interactions are consistently described by the ZBL potential. From Figure \ref{fig:single_defects_energy}b, a pronounced power-law relationship exists between the number of defects and PKA energy across all compositions. Hence, we employ a power function to fit the data, $N_d = AE^m$, where $N_d$ represents the number of defects, $E$ is the kinetic energy of the PKA, and $A$ and $m$ are material constants which were found weekly dependent on temperature in metals \cite{bacon2000primary}. $A$ is particularly affected by the definition of energy used \cite{bacon2000primary}, e.g., here, the electronic energy loss of the PKA is not considered and only the energy from elastic collisions is included. Instead, $m$ can serve as a generic variable to compare across different materials. Table \ref{tab:table_exp} summarizes the exponents for Al$_x$Ga$_{1-x}$N and literature values for several other material systems (belonging to metal, intermetallic, and ceramic) derived from MD simulations. In Al$_x$Ga$_{1-x}$N, the $m$ values are overall similar across different compositions while high-Ga systems tend to be slightly larger; they are also higher than those observed in metallic systems, which is attributed to the bonding characteristics of AlGaN. 

\begin{table}[!ht]
\centering
\caption{Summary of exponent ($m$) in the power function for different materials.}
\label{tab:table_exp}
\begin{tabular}{|l|l|l|}
\hline
Material                & $m$  & Source    \\ \hline
GaN                     & 0.95 & This work \\ \hline
Al$_{0.25}$Ga$_{0.75}$N & 0.96 & This work \\ \hline
Al$_{0.5}$Ga$_{0.5}$N   & 0.90 & This work \\ \hline
Al$_{0.75}$Ga$_{0.25}$N & 0.92 & This work \\ \hline
AlN                     & 0.90 & This work  \\ \hline
Al                      & 0.83 & \cite{bacon2000primary} \\ \hline
Ni$_3$Al                & 0.71 & \cite{bacon2000primary}  \\ \hline
Th$_{0.5}$U$_{0.5}$O$_2$              & 0.897 & \cite{jin2020systematic}   \\ \hline
\end{tabular}
\end{table}

\begin{figure}[ht]
    \centering
    \includegraphics[width=0.95\textwidth]{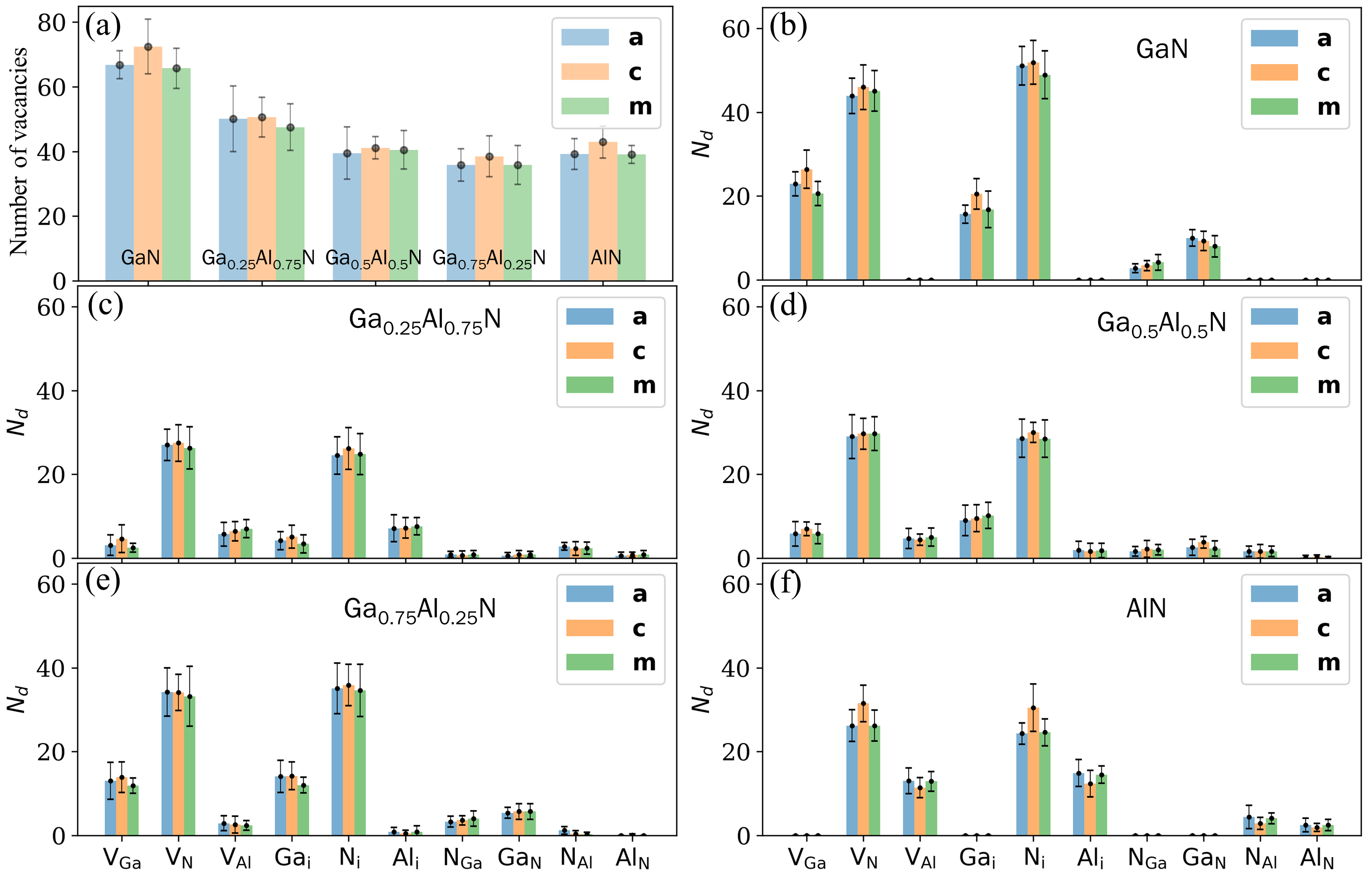}
    \caption{With 10 keV Ga PKA induced damage cascades initiated along $\mathbf{a}$, $\mathbf{c}$, and $\mathbf{m}$ directions in 300 K Al$_x$Ga$_{1-x}$N systems: (a) shows the number of vacancies, and (b-f) shows the number of individual defects.}
    \label{fig:single_defects}
\end{figure}

Next, we focus on the defects formed from the primary damage. Figure \ref{fig:single_defects} summarizes the speciation of point defects arising from the 10 keV cascade simulations presented in Figure \ref{fig:single_defects_energy}a, identifying ten types of point defects in the five compositions. Results deduced from PKAs along various crystalline directions ($\mathbf{a}$, $\mathbf{c}$, and $\mathbf{m}$) are distinguished. Notably, they do not show obvious differences in defect production. Previous experimental work with non-polar $\mathbf{a}$-plane and polar $\mathbf{c}$-plane epitaxial GaN samples \cite{lorenz2017implantation,mendes2019measuring} has shown distinct radiation responses in terms of defect morphology and population. Our current results suggest that these observed differences are likely not due to initial defect formation; instead, the differences may be attributed to the perpendicular strain-induced variations in defect migration. Consistent with Figure \ref{fig:single_defects_energy}a, Figure \ref{fig:single_defects}a indicates the total number of defects decreases with increasing Al content. N defects are predominant, which is expected, given that N has a lower threshold displacement energy compared to Ga \cite{hauck2024atomic}. Across all Al-containing compositions, the number of N interstitials is comparable to N vacancies; in GaN, N interstitials slightly outnumber N vacancies. Since N interstitials can capture electrons, acting as additional acceptors and leading to a positive shift in the threshold voltage of HEMT devices \cite{puzyrev2011radiation}, the prevalence of N interstitials could therefore have a strong impact on the properties of these devices. The total number of cation defects (Ga and Al) is much lower than that of N defects. Furthermore, there is a higher tendency to form Ga defects compared to Al defects, which is particularly exemplified in Al$_{0.5}$Ga$_{0.5}$N. Anti-site defects are generally much less common than other types of defects. This aligns with previous quantum mechanical calculations for recoil energies in the tens of electron volts range, which indicated the probability of generating an N antisite is lower than that of generating either a Ga or N vacancy \cite{puzyrev2011radiation}. When comparing GaN and AlN, it appears that within the anti-site defect category, $\mathrm{Ga_N}$ defects are more prevalent than $\mathrm{N_{Ga}}$, while $\mathrm{N_{Al}}$ and $\mathrm{Al_N}$ defects are similar and less frequent.


\begin{figure}[ht]
    \centering
    \includegraphics[width=0.95\textwidth]{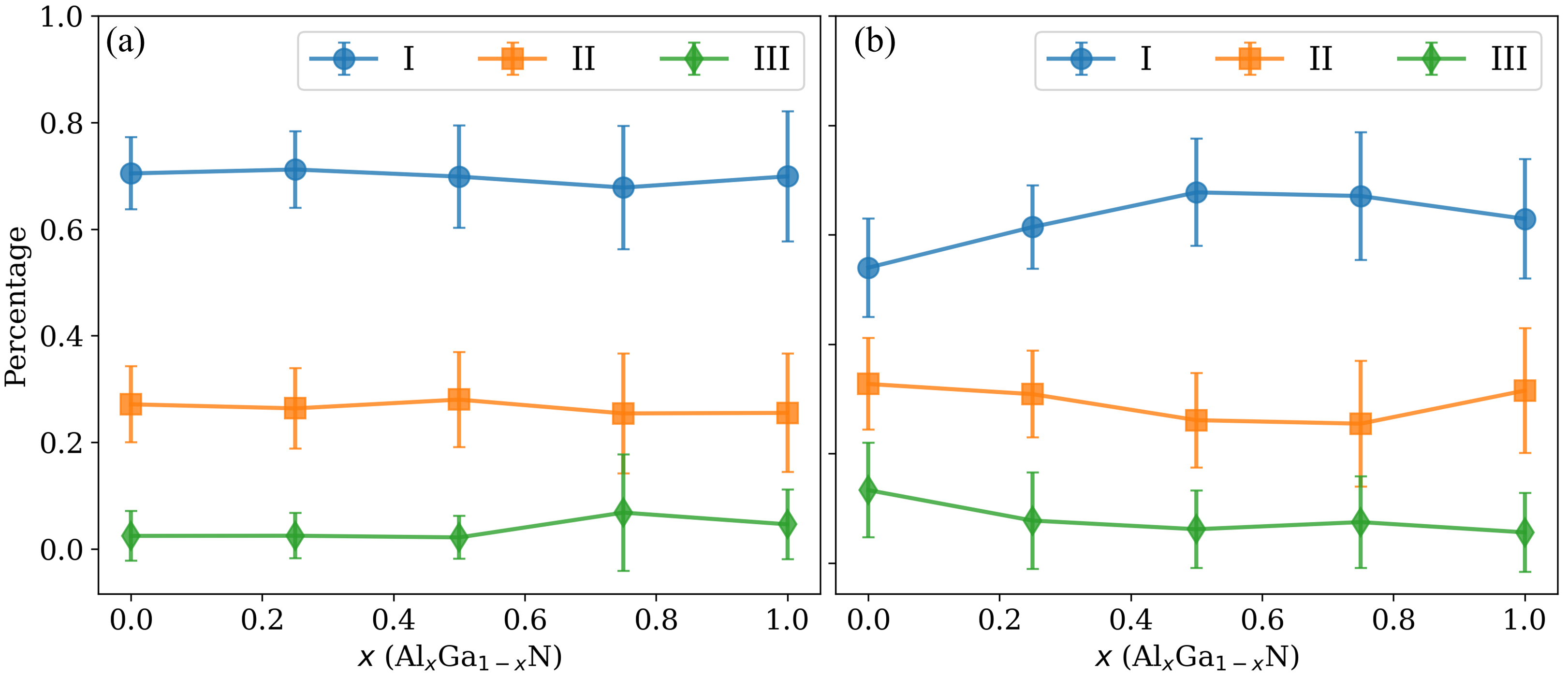}
    \caption{Fractions of interstitials (a) and vacancies (b) in the form of isolated point defect (I), di- and tri-cluster (II), and larger than size 3 clusters (III), at the end of 10 keV Ga PKA induced damage cascades in 300 K Al$_x$Ga$_{1-x}$N systems. }
    \label{fig:single_defect_clusters}
\end{figure}
 
We further examine the clustering characteristics of defects. In Figure \ref{fig:single_defect_clusters}, we have categorized the vacancy and interstitial defects into three groups based on the 10 keV damage cascade simulations: isolated point defects, di- and tri-clusters, and larger clusters (cluster size greater than three). It can be seen that the clustering behavior is weakly influenced by the composition. Approximately 70\% of interstitial defects are individually isolated ones, which is generally higher than the proportion of individual vacancies, particularly in GaN. Previous primary radiation damage studies on GaN also reported a higher fraction of individual interstitial defects compared to vacancies \cite{he2020primary}. This contrasts with simulation results in metals such as iron, where defects from primary damage tend to form clusters more frequently, with interstitials more likely to cluster than vacancies \cite{stoller1997primary}, which could be attributed to intra-cascade atom mobility. Here, we also observe that small defect clusters (di- and tri-clusters) constitute a significant portion, around 30\%, of both interstitials and vacancies. Larger clusters are much less common. However, in GaN, a notable fraction (13.3\%) of large vacancy clusters has been found. Such clustering behavior suggests that, when analyzing the trapping behavior of charge carriers, these small clusters are likely to have a significant impact on device performance under irradiation conditions, which necessitates further first principles evaluations.

\subsection{Simulation of Extended Radiation Damage}

With extended radiation damage simulations, we account for the interactions between current damage cascades and debris from previous cascades. The resulting defect morphology and population reflect the cumulative effects of radiation damage. Understanding these processes is essential for predicting the long-term behavior of materials under radiation exposure. Quantum mechanical calculations in GaN have shown that pre-existing defects can significantly lower the threshold energy required for defect generation compared to an otherwise perfect lattice ($>$20 eV versus 5-15 eV) \cite{puzyrev2011radiation}. Therefore, damage cascades interacting with pre-existing defect complexes can lead to a complex interplay between defect generation and annihilation, influencing the overall defect dynamics. This is relevant in high-dose radiation environments and in device layers that exhibit a high density of defects post-fabrication. 

\begin{figure}[ht]
    \centering
    \includegraphics[width=0.95\textwidth]{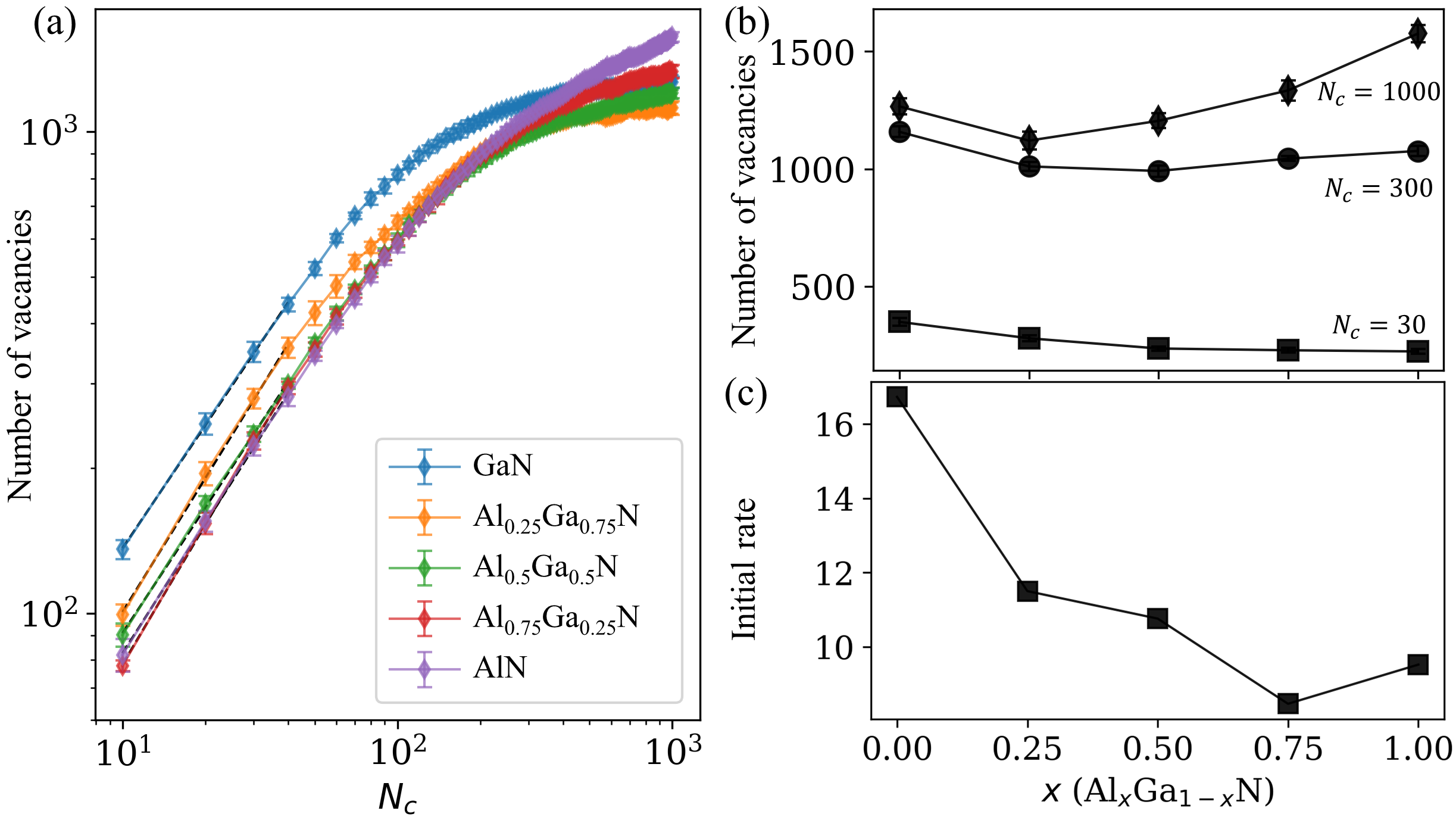}
    \caption{In 300 K Al$_x$Ga$_{1-x}$N systems: (a) Number of total vacancies with increasing irradiation; (b) Number of vacancies versus Al content at three selected doses ($N_c$ = 30, 300, and 1000); and (c) Initial defect accumulate rate versus Al content. Dashed lines in (a) are from the fitting to the power function.}
    \label{fig:multi_vacancies_300K_all}
\end{figure}

Figure \ref{fig:multi_vacancies_300K_all}a illustrates the total number of vacancies versu the number of cascades ($N_c$). The graph shows an initial rapid increase in vacancies, followed by a gradual saturation as the dose accumulates. This saturation behavior aligns with previous experimental findings in GaN, which indicated a gradual build-up of crystal disorder followed by saturation \cite{jiang2000situ,kucheyev2000damage,wang2002defect}. By comparison, the dose in transitioning to saturation regime appears earlier in GaN than in Al-containing alloys, also in line with experiments \cite{faye2016mechanisms}. Full amorphization at very high fluences has also been reported in experiments \cite{kucheyev2001effect}, although this is not addressed in the current modeling.

The behavior of defect accumulation is strongly dependent on the material composition. In GaN, the number of vacancies increases rapidly and reaches saturation earlier due to stronger annealing effects. In contrast, AlN shows a slower increase in vacancies, but the accumulation continues to higher defect concentrations by the end of the irradiation simulations. Such observations closely resemble experimental data on AlN and GaN. For example, during Eu implantation, AlN exhibits much lower defect levels up to medium fluences. However, at high fluences, AlN shows higher damage levels than AlGaN alloys and GaN \cite{faye2018crystal}. Strong agreement was also found with experimental work on Al$_{0.15}$Ga$_{0.85}$N and Al$_{0.77}$Ga$_{0.23}$N under Tm irradiation at different fluences, which shows that i) samples with 77\% Al approach the behavior of AlN and ii) higher damage in 15\% Al sample at lower fluences but lower damage at high fluence than the 77\% Al sample in the bulk sample region \cite{fialho2016impact}. Figure \ref{fig:multi_vacancies_300K_all}b compares the number of vacancies for different compositions across three doses. In the low dose case ($N_c$ = 30), GaN shows the highest damage level, while AlN has the lowest. As the dose increases to medium and high levels, the damage minimum shifts to lower Al content. This observation reflects experimental measurements obtained using RBS/C techniques \cite{faye2018crystal}. This indicates that alloying GaN with Al initially enhances defect recombination, reducing the overall defect population. However, as the Al content increases further, the defect accumulation increases at high doses. This increase is attributed to significant dislocation formation, which will be discussed later.

Additionally, it can be seen that during the initial stage of defect accumulation, where there is no strong overlap between cascades, the relationship between the number of defects and the number of cascades exhibits a linear trend on a log-log scale. This observation allows for fitting a power function to derive the defect accumulation rate in the low dose limit. Assuming the relationship $N = cN_c^\alpha$, where $N$ is the number of defects, $c$ is a scaling constant, and $\alpha$ is the exponent, we can calculate the initial rate of defect accumulation per cascade as $dN/dN_c|_{N_c=1}=c\alpha$. These calculated values are summarized in Figure \ref{fig:multi_vacancies_300K_all}c, which indicates that defect accumulation is fastest in GaN at low doses and generally decreases with increasing Al content. The sudden decrease from GaN to alloys highlights the enhanced defect recombination in low Al-containing alloys. However, as the Al content continues to increase, the gain in reducing the defect accumulation rate slows down. 

\begin{figure}[!ht]
    \centering
    \includegraphics[width=0.95\textwidth]{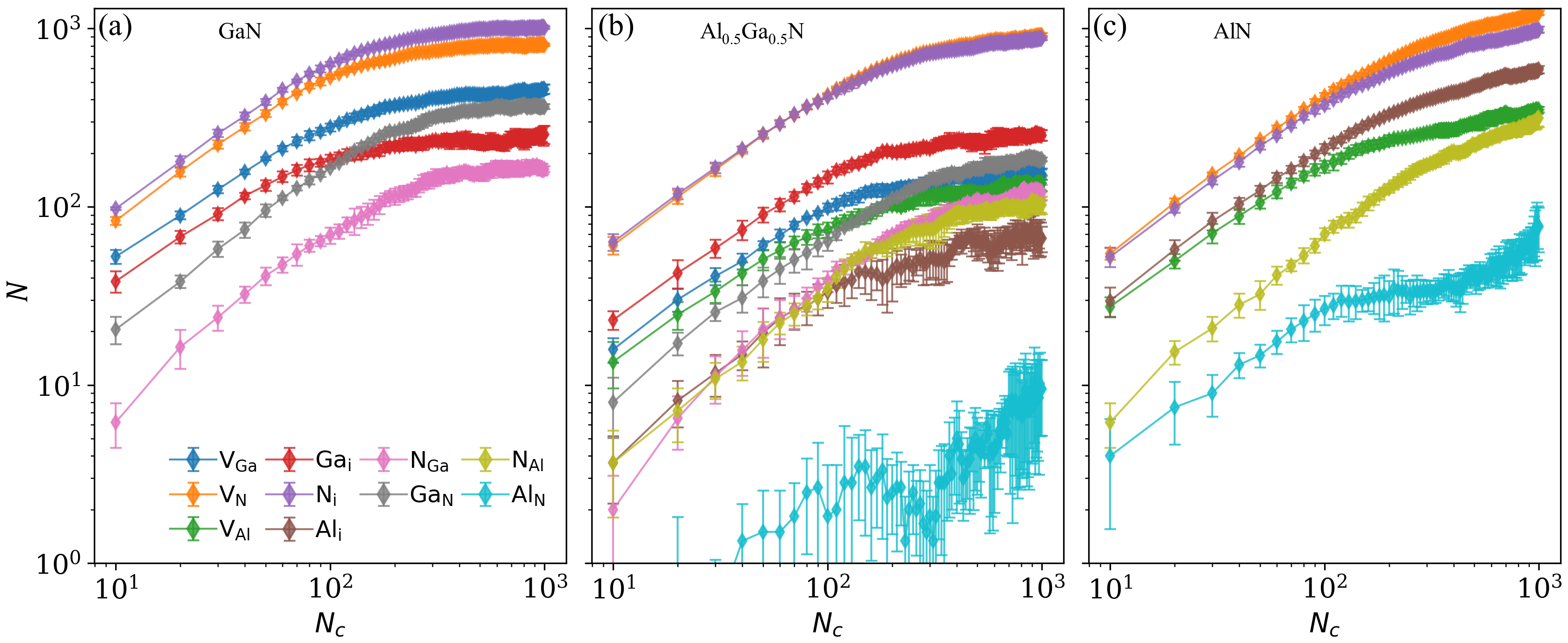}
    \caption{(a-c) Number of different defects with increasing irradiation in 300 K GaN, Al$_{0.5}$Ga$_{0.5}$N, and AlN, respectively.}
    \label{fig:multiple_defects}
\end{figure}
The variety of defects is separately quantified, as shown in Figure \ref{fig:multiple_defects} for GaN, Al$_{0.5}$Ga$_{0.5}$N, and AlN, respectively. Additional plots for Al$_{0.25}$Ga$_{0.75}$N and Al$_{0.75}$Ga$_{0.25}$N are provided in the SM. These figures show that nearly all types of defects initially increase in number and then gradually level off. However, the accumulation rates vary significantly across different defect types and compositions. N defects are predominantly observed, aligning with measurements on neutron-irradiated GaN samples \cite{kuriyama2006lattice}. In GaN, $\mathrm{N_i}$ outnumbers $\mathrm{V_N}$; in Al$_{0.5}$Ga$_{0.5}$N, $\mathrm{V_N}$ and $\mathrm{N_i}$ are comparable; but in AlN, $\mathrm{V_N}$ exceeds $\mathrm{N_i}$. In GaN, $\mathrm{V_{Ga}}$ increases at a higher rate than $\mathrm{Ga_i}$, resulting in a higher ultimate concentration of $\mathrm{V_{Ga}}$; among antisites, $\mathrm{Ga_N}$ increases drastically compared to $\mathrm{N_{Ga}}$, and eventually surpasses $\mathrm{Ga_i}$. In AlN, there is a noticeable trend of producing more $\mathrm{Al_i}$ than $\mathrm{V_{Al}}$, and $\mathrm{N_{Al}}$ defects are significantly more prevalent than $\mathrm{Al_{N}}$.  From the analysis of Al$_{0.5}$Ga$_{0.5}$N, we observe that the production rates of $\mathrm{V_{Ga}}$ and $\mathrm{V_{Al}}$ are close while $\mathrm{Ga_i}$ is substantially higher than $\mathrm{Al_i}$. Notably, $\mathrm{Ga_i}$ outnumbers $\mathrm{V_{Ga}}$, which is opposite to the trend in GaN. Similarly, $\mathrm{V_{Al}}$ exceeds $\mathrm{Al_i}$, opposite to the trend in AlN. For antisite defects, $\mathrm{N_{Ga}}$ and $\mathrm{N_{Al}}$ occur at similar rates, whereas $\mathrm{Al_N}$ is much less common compared to $\mathrm{Ga_N}$. The drastic differences between the alloy and the pure nitrides suggest a complex alloying effect in mediating defect accumulation. 

\begin{figure}[ht]
    \centering
    \includegraphics[width=0.95\textwidth]{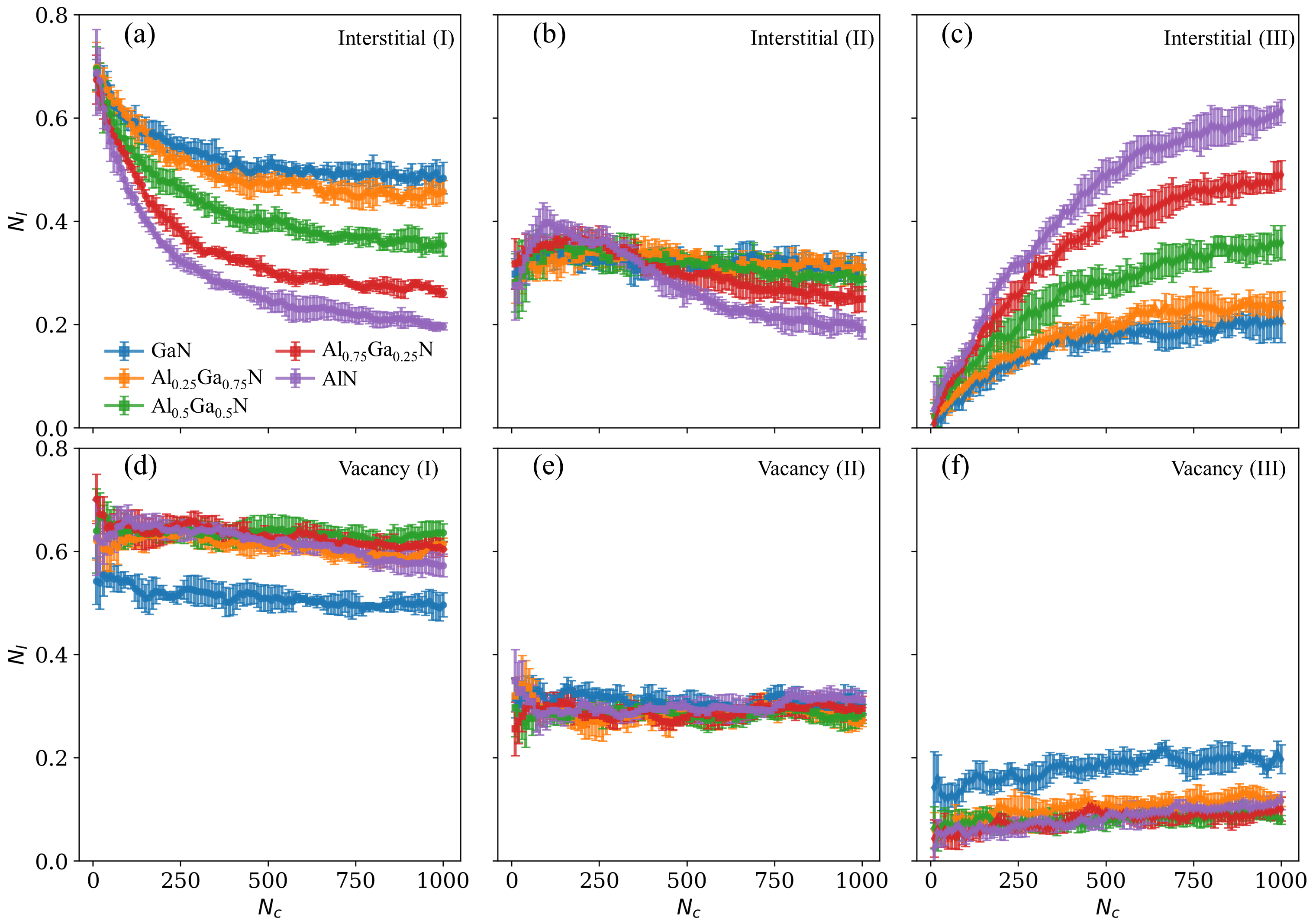}
    \caption{Fractions of interstitials in the form of isolated point defect (I), di- and tri-cluster (II), and larger than size 3 clusters: point defect (a), di- and tri-cluster (b), and larger than size 3 clusters (c), with increasing irradiation at 300 K, respectively. Fractions of vacancies in the form of isolated point defect (I), di- and tri-cluster (II), and larger than size 3 clusters: point defect (d), di- and tri-cluster (e), and larger than size 3 clusters (f), with increasing irradiation at 300 K, respectively.}
    \label{fig:multiple_defect_clusters}
\end{figure}
Similar to the single cascade simulations, we also examine the clustering behavior to understand defect accumulation mechanisms. Figure \ref{fig:multiple_defect_clusters} shows the clustering of interstitials and vacancies across the three defined categories, changing with radiation dose. From Figure \ref{fig:multiple_defect_clusters}a, it can be seen that the proportion of isolated interstitials (I) decreases with increasing dose, and gradually stabilizes. This reduction is notably more pronounced in systems with higher Al content; for example, during the saturation stage, AlN displays a significantly reduced fraction of isolated point defects at about 20\%, compared to GaN, which maintains around 50\%.  For di- and tri-interstitial clusters (II) as shown in Figure \ref{fig:multiple_defect_clusters}b, the trends vary with Al content. In systems with low Al content, there is a slight decrease in the proportion of these clusters across a wide range of dose. However, in systems with high Al content, there is an initial increase followed by a decrease in the proportion. This pattern indicates a dynamic balance between defect formation and annihilation within this defect category. In contrast, the fraction of larger interstitial clusters (size greater than 3, III) in Figure \ref{fig:multiple_defect_clusters}c consistently increases with dose, and this growth accelerates in systems with high Al content. For vacancy clustering, as shown in Figure \ref{fig:multiple_defect_clusters}d-f, the changes are much weaker, which is attributed to the lower mobility. The fractions of each category remain relatively stable, with weak increasing/decreasing trends as the radiation dose increases. Notably, for all the Al-containing systems, the fractions are similar. However, in GaN, the proportion of isolated vacancies is smaller, and the proportion of larger vacancy clusters (size greater than 3, III) is higher, suggesting a greater tendency to form large vacancy clusters in GaN than the alloys.
\begin{figure}[!ht]
    \centering
    \includegraphics[width=0.75\textwidth]{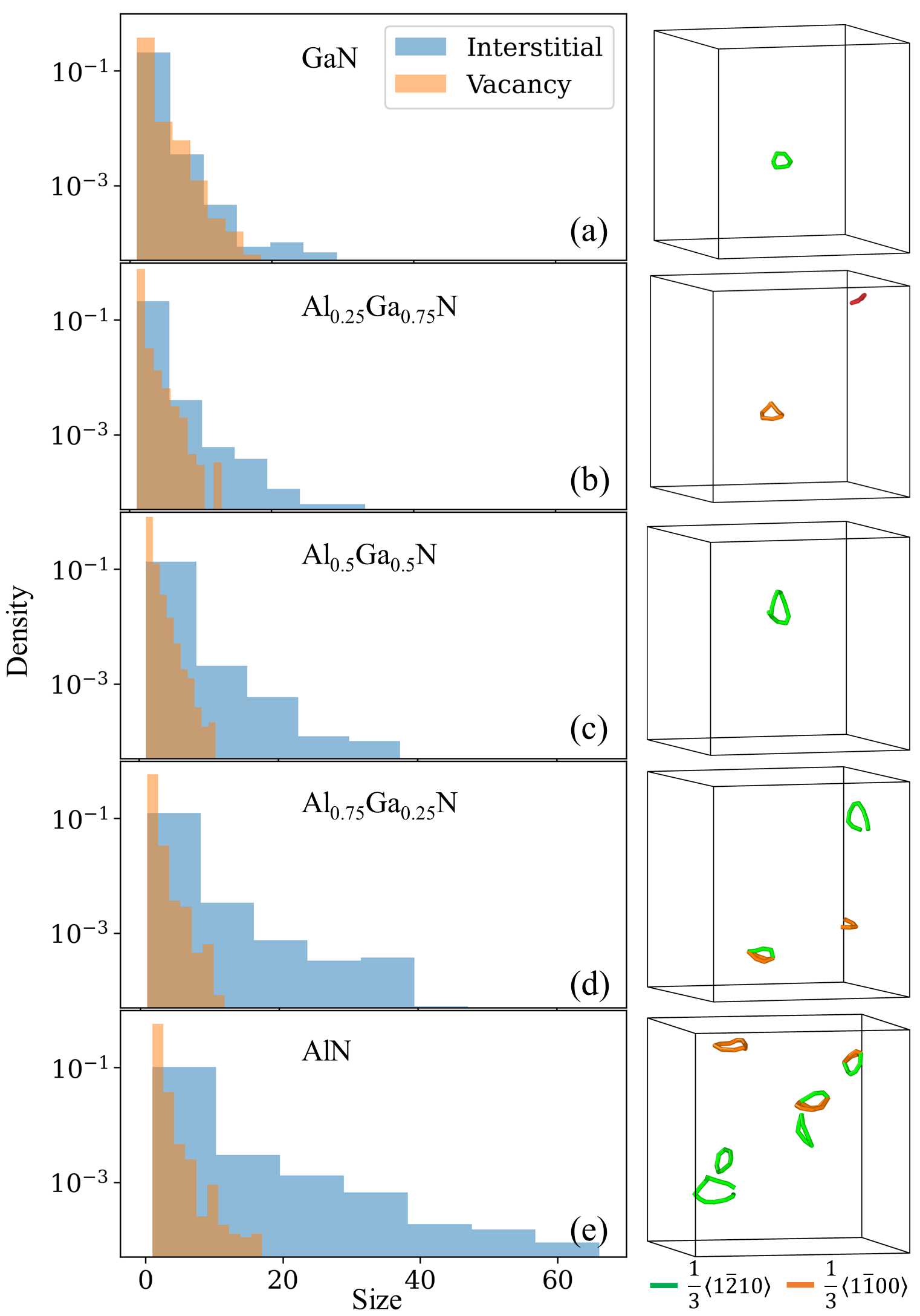}
    \caption{(a-e) Defect cluster size distribution at the end of irradiation simulations in 300 K Al$_x$Ga$_{1-x}$N systems. The right column indicates the corresponding atomic configurations containing dislocations.}
    \label{fig:multiple_defect_clusters_dist}
\end{figure}

To gain more insights into the defect clusters that dominate the defect evolution, we look into the atomic configurations. Figure \ref{fig:multiple_defect_clusters_dist} presents the size distribution of interstitial and vacancy clusters, along with dislocations in the structure. Overall, vacancy clustering is significantly more limited than interstitial clustering, consistent with Figure \ref{fig:multiple_defect_clusters}. With increasing Al content, there is a pronounced tendency for interstitial clustering, particularly the enhanced formation of dislocation loops. In GaN, prismatic dislocation loops are observed in simulations. Multiple experimental studies on irradiated GaN also indicate strong defect annealing, which promotes enhanced defect recombination and the formation of extended defects such as stacking faults and dislocation loops \cite{kucheyev2001ion,wendler2013comparison,lorenz2017implantation}. These studies have noted that planar defects parallel to the basal plane are primarily found in GaN thin films under ion irradiation, even at liquid nitrogen temperatures \cite{kucheyev2000damage,wang2002defect}. Although Wendler et al. attributed the formation of extended defects at cryogenic temperatures to damage-induced strain \cite{wendler2013comparison}, our results suggest that strain may not be the primary factor in forming extended defects, but the cumulative/overlapping damage events could play a major role. Different microstructural responses were also noted for $\mathbf{a}$-plane oriented GaN versus $\mathbf{c}$-plane oriented GaN where dominant extended defects in $\mathbf{c}$-plane GaN are basal stacking faults while dislocation loops are more commonly formed in $\mathbf{a}$-plane GaN \cite{catarino2012enhanced,lorenz2017implantation}.  In contrast, our current simulations reveal that the loop habit planes are not confined to the basal plane but manifest across various crystallographic planes. This discrepancy is likely due to the strain states in epitaxial thin film samples in experiments, where radiation-induced stress can relax along the free surface's normal direction while the lateral directions remain constrained by the substrate. This sample configuration can influence defect accumulation during irradiation and likely lead to the preferential condensation of interstitials on the basal plane. Our current modeling strategy does not capture the effects of macro strain on defect evolution. The observed differences in the orientation of planar defects between simulations and experiments suggest that strain effects could be key to the different microstructural features.  

In AlN, our simulations indicate a much higher density of loops compared to GaN by the end of the radiation simulations. During irradiation, these mobile loops enhance interactions with defects generated from damage cascades, potentially acting as precursors to larger defect complexes observed under transmission electron microscopy (TEM) \cite{ruterana2013mechanisms}. Experimental work on irradiated epitaxial AlN thin films reports both prismatic and basal stacking faults, with a prominent formation of basal stacking faults \cite{leclerc2012mechanisms,jublot2019temperature,ruterana2013mechanisms}. Under He irradiation, the formation of basal stacking faults was hypothesized to be due to He bubbles, which promote the nucleation of these faults and point defect clusters \cite{jublot2019temperature}. Studies involving Eu irradiation suggest that the high damage observed in AlN at high fluences could be due to nitrogen deficiency leading to local lattice distortions \cite{faye2018crystal}. However, our current simulations provide an alternative explanation, showing a pronounced tendency to form defect clusters, particularly interstitial dislocation loops, in the absence of noble gas bubbles and nitrogen deficiency. The distinct damage formation and the cumulative effects of overlapping cascades between GaN and AlN is proposed to be the reason for the microstructural differences at high fluences.

\begin{figure}[ht]
    \centering
    \includegraphics[width=0.75\textwidth]{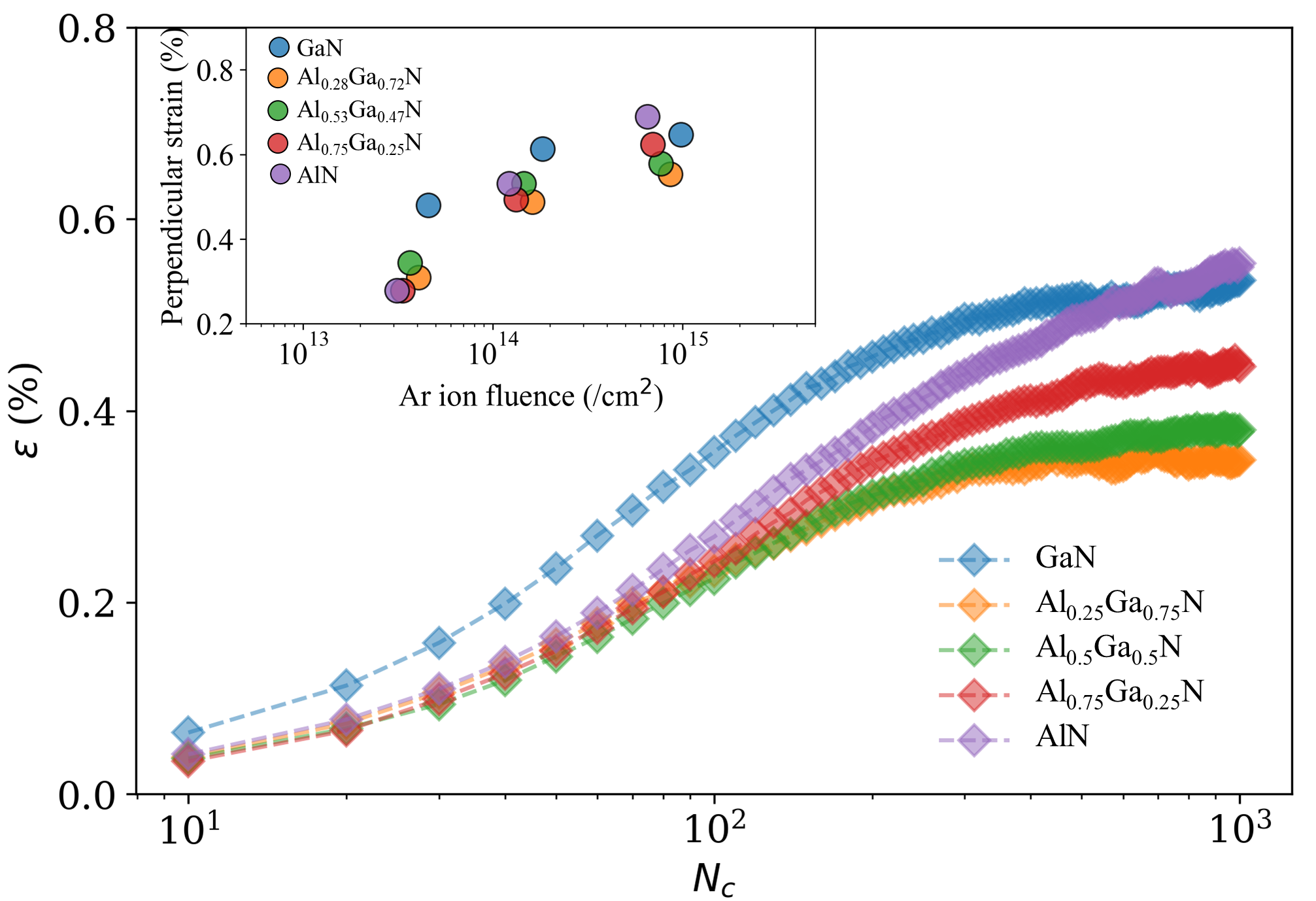}
    \caption{Average normal strain ($\epsilon$) with increasing irradiation in 300 K Al$_x$Ga$_{1-x}$N systems. The inset shows the experimental measurement of perpendicular strain in Al$_x$Ga$_{1-x}$N under Ar ion irradiation at room temperature from \cite{faye2016mechanisms} (adapted with permission).}
    \label{fig:multiple_strain}
\end{figure}

Lastly, we examine radiation-induced strain, a quantity frequently reported in irradiated AlGaN samples, allowing us to validate the calculations to some extent. Strain evolution as a function of fluence can be directly quantified through X-ray diffraction (XRD) techniques, which measure small variations in lattice parameters. With epitaxial samples on a substrate used in radiation experiments, it is intuitive that the lattice can expand perpendicular to the surface, while in-plane strain is constrained by the substrate \cite{mendes2019measuring,lorenz2017implantation}. Indeed, it has long been reported that ion implantation leads to uniaxial strain in epitaxial thin films, which increases with fluence \cite{wendler2013comparison,lorenz2017implantation,sharma2021effect}. For example, a study involving GaN/Al$_2$O$_3$ samples irradiated with 92 MeV $^{129}$Xe$^{23+}$ at various fluences found that irradiation caused lattice expansion along the $\mathbf{c}$-direction (parallel to the ion path) while showing minimal variation along the $\mathbf{a}$-direction (perpendicular to the ion beam) \cite{eve2022structural}. Since i) the strain effect was not considered during the cascade simulations and ii) the initial PKA directions were random instead of preferential forward scattering as expected in ion irradiation \cite{shao2014effect}, we alternatively relax the atomic configurations as a function of dose under the isothermal-isobaric ensemble at 300 K and zero external pressure. Then, the non-zero strain perpendicular to the in-plane is estimated based on the average normal strain. Figure \ref{fig:multiple_strain} shows the average normal strain values, with experimental measurements by Faye et al. \cite{faye2016mechanisms} included in the inset for comparison. Detailed calculation results for the strain along the individual crystalline directions ($\mathbf{a}$, $\mathbf{m}$, and $\mathbf{c}$) are provided in the SM. For all compositions, the lattice expansion with increasing dose exhibits a rapid increase followed by saturation, mirroring the defect population trends shown in Figure \ref{fig:multi_vacancies_300K_all}. Notably, these calculated strain profiles compare reasonably well with experimental data and trends of strain evolution with increasing ion fluence, based on Al$x$Ga$_{1-x}$N epitaxial films irradiated with Ar ions at room temperature \cite{faye2016mechanisms}. We note consistent observations: i) GaN generally shows higher strain compared to Al-containing alloys; ii) it appears that 25\% Al leads to relatively smaller strain values, and iii) while AlN initially exhibits lower strain than GaN, it eventually surpasses GaN as the dose increases. It is important to note that strain values can vary spatially in ion-irradiated samples due to defect dynamics affected by surface, depth-dependent defect production, and chemical effects from the implantation \cite{jublot2019temperature,fialho2018defect}, so the comparison presented here should be perceived more qualitative. For instance, in He ion-irradiated AlN epitaxial films at room temperature, there is a notable difference between the near-surface and deeper layers, with the elastic strain  0.5\% near the surface and 2\% in the deeper layer. Radiation types could also lead to differences, for example, GaN samples implanted to $1\times10^{15}$ Ar ion/cm$^2$ (in the saturation fluence range) show measured strain values of approximately 0.7\%-0.75\% in GaN perpendicular to the sample growth plane \cite{wendler2013comparison,mendes2019measuring}, while in neutron-irradiated epitaxial GaN layers (fluence at $8\times 10^{19}$ cm$^{-2}$), the $\mathbf{c}$-axis lattice constant increases by 0.38\% \cite{boyko2011effect}. In epitaxial AlN thin films irradiated with He ions, a uniaxial $\mathbf{c}$-axis tensile strain of 1.0\% was reported at the highest fluence considered ($1\times 10^{17}/$cm$^2$) \cite{sharma2021effect}. The current calculated values fall within the measured range of strain values, suggesting that the extent of damage is likely comparable to that observed in experimental work.

\section{Discussions}

Radiation-induced crystalline defects represent a diverse array of structural imperfections compared to thermal defects, which are primarily point defects due to their relatively low formation energies. Numerous experimental data have demonstrated that MeV-level ion radiation leads to the gradual degradation of device performance. Unlike single-event effects, the prolonged crystalline damage and resultant performance decline suggest that microstructural defect clusters play an important role in this process. These newly formed defect types are expected to trap charge carriers, though further investigation is ongoing to understand their charge transition levels and carrier capture cross-sections. Moreover, radiation-induced defects can enhance diffusion processes, and the extent of diffusion enhancement is directly proportional to defect concentration. Previous research by Kuball et al. \cite{kuball2011algan} suggested that diffusion is a critical factor in the initial stages of degradation in AlGaN/GaN HEMTs. The increased impurity diffusion was attributed to inverse piezoelectric stress and gate leakage currents, potentially leading to degradation mechanisms such as pit or groove formation \cite{kuball2011algan,chowdhury2008tem}. When considering radiation-induced defects, which result in a significantly higher defect concentration than thermal equilibrium conditions, this enhanced impurity diffusion becomes even more pronounced and contributes to device performance degradation.  

Ion beam-induced radiation damage has been extensively studied in GaN, while there has been significantly less work on AlGaN alloys. One important aspect to consider is the effect of temperature, which is directly relevant to the application environments of these materials. Experimentally, it has been observed that even at low temperatures, such as 15 K, significant formation of extended defects occurs in GaN \cite{lorenz2017implantation}. As the temperature increases, the damage fraction versus Ar ion fluence at room temperature and 15 K, determined using in situ RBS/C, suggests similar behavior at both temperatures \cite{wendler2013comparison}. At room temperature, there is only a slight shift in the damage profile due to enhanced defect kinetics. It hints that the thermal effect in driving defect evolution is not the primary factor up to room temperature. A similar conclusion was drawn based on the electrical characteristics of GaN implanted with Si up to 400 $^\mathrm{o}$C \cite{irokawa2005electrical}. Additionally, in AlN, increasing the irradiation temperature from liquid nitrogen temperature to room temperature has a relatively minor effect on the production of stable structural damage \cite{kucheyev2002ion}. Hence, as a validation, we also conduct simulations at 15 K, keeping all other settings the same. Our findings show that the differences between the 15 K and 300 K simulations are minor (see SM for defect calculations at 15 K). While reducing the temperature can affect the absolution values of the defect population, the overall trend remains consistent across all compositions. Furthermore, the defect accumulation trends in our simulations align well with RBS data for both GaN and AlN at 15 K \cite{wendler2016ion}. Hence, we conclude, in all these alloy materials under ion irradiation, that thermal effects play a minor role in contributing to the damage accumulation, at least up to room temperature, and the primary driver of microstructural evolution is radiation-induced damage cascades and cumulative effects. However, it is important to note that with even higher temperatures, thermal effects should be considered, e.g., TEM characterization shows suppression of basal stacking faults formation after Tb ion implantation in Al$_x$Ga$_{1-x}$N ($x$=0.20, 0.50, 0.63) films at 550 $\mathrm{^o}$C \cite{fialho2018defect}. At such high temperatures, thermally activated processes promote defect evolution, which the MD simulations cannot capture due to limitations in timescale. 



Up to high doses, experimental studies report multi-stage microstructural evolution \cite{lorenz2017implantation,kucheyev2002ion}, partly captured in the current simulations within the simulated dose range. Referring to 300 keV Ar irradiation data from \cite{wendler2016ion}, which share the same trend with the current results, GaN exhibits a high damage formation rate, especially at low fluences where cascade overlapping is not significant. However, at higher fluences, alloys with a high AlN fraction begin to exhibit greater damage, with AlN showing the highest damage levels. This increased damage in AlN at high fluences is confirmed to be due to the formation of extended defects, as hypothesized based on experimental studies using Eu ion implantation \cite{leclerc2012mechanisms}. Aligning with previous work arguing a lower damage production cross-section for AlN compared to GaN, we found the initial rate of damage accumulation is lower in Al-containing compounds than in GaN, and notably, alloying Al in GaN at 25\% could be overall beneficial in terms of radiation resistance, as it not only reduces the defect accumulate rate but also reduce the saturation defect density. No amorphization is observed in all systems within the dose range. Experimental data indicate that amorphization in these materials begins at much higher doses \cite{kucheyev2000damage,faye2018crystal,kucheyev2004dynamic}. Notably, AlN demonstrates superior resistance to radiation-induced amorphization compared to GaN \cite{kucheyev2002ion,ruterana2013mechanisms}. Even at cryogenic temperatures, the AlN lattice remains crystalline at very high doses \cite{kucheyev2002ion}. The radiation stability of all these compounds is attributed to efficient inter-cascade annealing, where defect complexes are identified from the single cascade simulations, and the overlapping of these cascades leads to the formation of dislocation loops and diverse defect clusters. Based on the current simulations, we delineate these dynamic annealing effects into more defined regimes: it is strong in GaN at high fluences, enabling early build-up and subsequent saturation of defects; in AlN, it is more pronounced at low fluences; in the alloy compounds, this annealing effect remains robust across a broad dose range. These results imply the complex interplay between defect formation, dynamic annealing, and cascade effects in determining the radiation tolerance of AlGaN materials.

 
Next, we discuss the limitations of the current modeling strategy and suggest areas for future improvement. The present approach models prolonged radiation damage using a constant PKA energy. In the SM, we show the PKA energy spectra resultant from 5 MeV iron and 2 MeV proton implantation into GaN, where the majority of PKAs are below 10 keV. We chose 2 keV for the PKA energy in the simulations. This value is a reasonable compromise: it is not too high to incur excessive computational costs due to the size of simulation cells, and not too low to fail to induce damage cascades. Hence, the simulations represent a radiation process driven by damage cascades, but future work could incorporate more realistic recoil energy spectra. This energy spectrum varies for different ions used in radiation experiments, leading to distinct outcomes due to inherent differences in energy transfer during collision processes. For example, it was shown that different ions could produce varying levels of damage and radiation-induced strain in Al$_x$Ga$_{1-x}$N at the same fluences (see a comparison of Ar, Xe, and Eu radiation data in \cite{faye2018crystal}). In addition, although significant similarities in damage buildup in GaN by heavy and light ions were noted by Kucheyev et al. \cite{kucheyev2002ion}, differences exist in the saturation ion fluence, chemical effects (especially near the peak damage region, which can suppress defect recovery due to solute trapping effects \cite{jiang2000situ,fialho2018defect}), and damage accumulation rates \cite{kucheyev2000damage}. In AlN, 85 MeV Fe irradiation led to an increase in the $c/a$ ratio and a reduction of AlN density with fluences up to $10^{14}$ ion/cm\(^2\) \cite{dukenbayev2019investigation}, whereas 229 MeV Xe ion irradiation at similar fluences resulted in negligible changes in the $c/a$ ratio \cite{kozlovskiy2018dynamics}. Therefore, in this work, we are not converting the number of cascades into fluence or dose for experimental comparison; instead, we use trends to delineate different dose regimes (defect increase and saturation stages) to provide a generic understanding. In addition, ion implantation experiments often exhibit strong surface effects due to the limited depth of implanted ions. The varying damage profile along the depth creates significant inhomogeneity in the microstructure, e.g., multi-layered microstructural features, and the measured strain \cite{kucheyev2000damage,faye2018crystal,fialho2018defect}. Given the limited size of simulation cells in MD simulations and the exclusion of surface effects, the agreement with literature data focusing on the bulk-like deep layers is satisfactory.

Finally, another aspect that could be improved in future work is the timescale disparity in the simulations. The use of consecutive damage cascades results in a much higher dose rate than that in experiments. While this approach is acceptable for comparing with low-temperature radiation experiments, where thermal effects are minimal, the discrepancy in timescale still introduces uncertainties when comparing microstructures, especially at elevated temperatures. Future improvements could include extending the timescale via time-accelerated methods such as hyperdynamics \cite{voter2002extending}. Then, the interval between the cascades can be largely increased to make the simulated dose rates more comparable to experimental dose rates, although it would come with increased computational costs. A related consideration is the radiation-induced strain. Our simulated radiation damage process does not consider the stress relaxation experienced in realistic epitaxial samples, where the surface direction is stress-free. While single cascade simulations for defect production are likely not to be significantly impacted by these very low strain values, this stress relaxation likely influences defect evolution by altering atom diffusion dynamics, especially diffusion anisotropy induced by small strain. Future work could incorporate strain effects in simulations along with time-accelerated methods. This would enable sufficient defect evolution within a strain field, providing a more accurate representation of defect dynamics to improve the reliability of simulation results in predicting real-world materials behavior under radiation.


\section{Summary}

In this work, we investigate the radiation-induced defect formation and evolution in Al$x$Ga$_{1-x}$N alloy compounds at the atomistic level, an area previously unexplored in detail. While extensive experimental results exist, particularly for GaN, understanding the underlying mechanisms of damage accumulation requires interpreting high-length-scale experimental data, which lacks support from an atomistic understanding of defect dynamics. Our simulations successfully reproduce several key experimental observations, such as earlier damage build-up in GaN compared to AlN, significant damage in AlN at high doses, reduced damage level in alloys, and reasonable quantitative agreement in radiation-induced strain. This validation allows us to probe the atomistic details of damage accumulation and reveal underlying mechanisms across the compositional spectrum. We find that alloy composition significantly modifies the potential energy landscape for defect dynamics, resulting in distinct patterns of defect generation and microstructural features. For instance, a high Al fraction results in less initial defect generation but increases defect concentration at higher doses. Notably, an optimal Al content ($\sim$ 25\% Al) for radiation resistance exists between the two ends of the alloys. The high damage levels in AlN at high doses are explained by a preference for large cluster formation and mobile dislocation loops, accelerating the formation of large defect complexes. Thus, this work fills in the gap of detailed atomistic insights into radiation-induced lattice damage in AlGaN, offering information for optimizing these materials for radiation performance.

\section*{Acknowledgments}
 
This work is supported by the Air Force Office of Scientific Research under award number FA9550-22-1-0308.  
 
 \bibliographystyle{elsarticle-num} 
 \bibliography{manuscript}





\end{document}